\documentclass[12pt]{aastex}
\usepackage{emulateapj5}

\def\cc{cm$^{-3}$}
\def\ccc{cm$^{-3}~$}
\def\apj{ApJ}
\def\aa{A\&A}
\def\aas{A\&AS}
\def\araa{ARA\&A}
\def\pasp{PASP}
\def\aj{AJ}
\def\mn{MNRAS}
\def\a{$\alpha$~}
\def\b{$\beta$~}
\def\lr{low-resolution~}
\def\brg{Br$\gamma$\/\ }

\def\h92{H92$\alpha$}

\def\ne{{n$_e$\/\ }}
\def\sn{{`supernebula'~}}
\def\nemin{{n$_e^{min}$\/\ }}
\def\l0{{l$_0$\/\ }}
\def\emu{{pc cm$^{-6}$\/\ }}
\def\tc{{$\tau_c$\/\ }}

\def\s{s$^{-1}$\/\ }
\def\nlyc{N$_{Lyc}$\/\ }
\def\nlycc{N$_{Lyc}$}
\def\asec{$^{\prime\prime}$\/\ }
\def\asecc{$^{\prime\prime}$}

\def\aminn{$^{\prime}$}
\def\deg{$^{\circ}$\/\ }
\def\msun{$M_{\odot}$\/\ }
\def\lsun{$L_{\odot}$\/\ }
\def\msunn{$M_{\odot}$}
\def\lsun{$L_{\odot}$\/\ }
\def\ha{H$\alpha$\/\ }
\def\hb{H$\beta$\/\ }
\def\kms{km s$^{-1}$}

\begin{document}
\title{VLA Observations of the H92$\alpha$ line from NGC 5253
 and Henize 2-10 : Ionized Gas around Super Star Clusters}
\author{Niruj R. Mohan}
\affil{Raman Research Institute, Bangalore, India}
\affil{and}
\affil{Indian Institute of Science, Bangalore, India}
\email{niruj@rri.res.in}
\and
\author{K.R. Anantharamaiah}
\affil{Raman Research Institute, Bangalore, India}
\email{anantha@rri.res.in}
\and
\author{W.M. Goss}
\affil{National Radio Astronomy Observatory, Socorro, USA}
\email{mgoss@aoc.nrao.edu}
\shorttitle{RRLs from NGC 5253 and He 2-10: Ionized gas around SSCs}
\shortauthors{Mohan, Anantharamaiah and Goss}

\begin{abstract}

We have detected the \h92 radio recombination line from two dwarf starburst galaxies, NGC 5253 and
He 2-10, using the Very Large Array. Both the line data as well as the radio continuum data are used to 
model the properties of the ionized gas in the centers of these galaxies. 
We consider a multi-density model for radio recombination lines and show why
previous models, which were based on the assumption of gas at a single
density, are valid in many situations. The models show that the ionized gas
has a density of $\sim$10$^4$ \ccc in both galaxies, with an effective size of 2-10 pc and a total mass
of $\sim$10$^4$ \msunn. The derived production rate of
Lyman continuum photons is $\sim 2.5\times 10^{52}$ s$^{-1}$ in both the
galaxies and the corresponding mass of stars (assuming a Salpeter IMF) is
$\sim10^5$ \msunn.  The implied stellar density shows that the observed
radio recombination lines arise from ionized gas around super star
clusters (SSCs) in both galaxies (these SSCs have been recently detected
through their radio continuum emission).
The existence of $\sim$10$^4$ \msun of ionized gas within a few parsecs of an 
SSC places strict constraints on dynamical models. Using simple arguments, the parameter
space for a few possible models are derived. The well known radio-FIR
correlation also holds for NGC 5253, although the radio emission from this
galaxy is almost completely thermal. It is shown that NGC 5253 is strong
evidence that the component of FIR emission from warm dust is correlated
separately with the component of radio emission from thermal
bremsstrahlung.

\end{abstract}


\section{Introduction}

It is now known that star formation in starburst galaxies is not just a
scaled-up version of quiescent star formation in normal galaxies. For example, the phenomenon of
starbursts involves the formation of super star clusters
(SSCs), which are believed to be the pre-cursors of the present-day
globular clusters (see recent review by \citet{whi00} and references therein). These SSCs 
may generate large-scale
superwind outflows \citep{ham90} which can carry away a good
fraction of the gas and processed heavy elements from the starburst region.
Star formation also occurs in the form of individual field stars
and small unresolved clusters. Along with stellar emission diagnostics, studies of 
ionized gas around OB stars provide information on the young stellar population and also 
characterizes the surrounding interstellar medium. Radio continuum 
and recombination line emissions are good extinction-free probes of this gas. 
High resolution radio continuum imaging of 
starburst centers, for example, has led to the discovery of a host of pc-sized
thermal and non-thermal compact sources (NGC 253: \citet{ua97} and references therein; 
NGC 2146: \citet{tar00}; M 82: \citet{kbs85}; NGC 4038/9: \citet{nul00}). 
Studying the compact thermal sources is difficult in the UV, 
optical or even in IR, owing to extinction problems. A nebular diagnostic like 
radio recombination 
lines (RRLs) can probe thermal gas with large emission measures.
In the cm-wavelength range, RRLs have been shown to be
detectable from nuclear starburst regions in a number of nearby galaxies 
\citep{scr77,sb97,pbm91,azgv93,zagv96,pag98}.
These lines have proven to be good probes of gas density, and are reliable
estimators of ionized gas mass and the ionizing photon flux in these starburst regions. 
RRLs have been studied in about 14 galaxies so far (\citet{anan00} and references
therein). 
In this paper we present observations of the \h92 RRL at 8.3 GHz 
from two prototypical Blue Compact Dwarf Galaxies (BCDGs), 
NGC 5253 and He 2-10. 

BCDGs are dwarf galaxies which are dominated by an ongoing starburst, with or without a
background quiescent star formation. NGC 5253 and He 2-10 are classic examples \citep{dsw98}
of BCDGs. NGC 5253 (D=4.1 Mpc, Saha et al. 1995) \nocite{saha95} is a 
dwarf galaxy with an 
underlying elliptical structure \citep{caldp89}. This galaxy has large 
scale ionized filaments \citep{graham81} and a 
number of young star
clusters in the halo \citep{vdb80}. The central kpc region is undergoing active star
formation and hosts a number of SSCs \citep{tbh00,calzetti97,gorjian96,beck96,meurer95,caldp89}.
Wolf-Rayet emission has been detected in the central region \citep{camp86,walshr87,walshr89,schae97}.
Based on modeling of the stellar emission and through the presence of ambient ionized gas and 
Wolf-Rayet features \citep{gorjian96}, a spatial age gradient in the star formation in the
core has been inferred. NGC 5253 is notable for having a flat radio
continuum spectrum \citep{leq71,beck96}, which is attributed 
to lack of non-thermal radiation due to the absence of supernova remnants. This absence is 
believed to be due to the very young age of the star formation
activity. The resulting age estimate is consistent with
that derived from optical and IR spectrophotometry \citep{beck96}. The 
\ha structure in the
inner 20\asec is complex and there is evidence for a patchy dust
distribution (Calzetti et al. 1997; hereafter CMB97) \nocite{calzetti97}.
Recently, \citet{tbh00}, hereafter TBH00, have made a high resolution image
of the central region in the radio continuum at 15 and 22 GHz and have detected a compact 
source which they interpret as a \sn surrounding a nascent SSC. 

He 2-10 (D=9 Mpc) was the first Wolf-Rayet galaxy discovered \citep{awg76}. 
Three star forming regions have been identified in the central part of the galaxy and among them, 
the central $\sim$5\asec region, called region `A', is the youngest \citep{mendez99}.
This region exhibits predominant 
Wolf-Rayet spectral features and is also the most active star forming region. HST imaging 
in the UV has 
shown the presence of a series of young clusters, believed to be SSCs \citep{cv94,john00}.
Deep \ha images of this galaxy \citep{mendez99} indicate a large scale bipolar outflow of
gas. Molecular gas, imaged through CO lines, shows a disturbed morphology with a rotating disk or a bar 
\citep{kobul95}. Radio continuum from He 2-10 has an overall spectral index of 
$-$0.6 and is therefore dominated by non-thermal emission \citep{awg76}. 
Recent high resolution radio continuum imaging by \citet{kj99}, hereafter KJ99, 
has shown the existence of five compact radio
knots with flat spectra. These radio knots are interpreted by the authors as high
density nebulae surrounding young SSCs, which are probably totally obscured 
in the optical. 

Based on their nuclear \brg flux densities and the presence
of active star formation in the central region, 
we selected these two galaxies to search for RRLs from the central few
hundred parsecs. Our observations are a part of a program we
have undertaken to detect RRLs in nearby starbursts and to follow up the
detected sources using multi-frequency RRL and continuum observations. 
The aim is to model the density structure, kinematics and geometry of the ionized gas 
in these regions and to determine an extinction-independent star formation rate. 
An example of such a study of Arp 220 has been presented by \citet{anan00}. 

\section{Observations and Results}

NGC 5253 and He 2-10 were observed with the Very Large Array (VLA) in both the CnB and the DnC
configurations with an on-source integration time of 3 hours per source in each configuration.
Some details of the observations are given in Table 1. 
The system was tuned to detect the \h92 line ($\nu_{rest}$=8309.385 
MHz) using a total bandwidth of 25 MHz and 15 spectral channels. Data from both
polarization channels were recorded. The data was hanning-smoothed offline to reduce 
the effects of Gibbs ringing. The effective velocity resolution is 113 \kms. 
The flux density scale was fixed by using 3C 286 as a calibrator and phase and bandpass
calibrations were done by using a nearby point source. All the data analysis were
done by using standard algorithms within the software AIPS. The continuum visibility data 
was self-calibrated and the resultant antenna gains were applied 
to the line data as well. The CnB and DnC array data 
of each galaxy were later concatenated and combined (C+D) continuum and line 
datasets (with a resolution of $\sim$5\asecc) were obtained. The final line images 
were derived after subtracting the continuum using the AIPS task 
UVLIN which subtracts a linear function of frequency from the real and imaginary 
parts of the observed visibilities \citep{cornwell92}. 
The continuum and line data are presented in Table 2. 
Figure 1 shows the continuum images of the two galaxies. The images show evidence for
filaments, especially in the case of NGC 5253, where there are extensions 
in the southern direction. These filaments are also seen in optical line 
and continuum images and extend out to a few arcminutes. 
The total continuum flux density of NGC 5253 at 8.3 GHz is 58 mJy. The 
radio spectrum of this galaxy is known to 
be flat from 1.4 GHz up to 23 GHz, based on both single-dish and scaled array VLA observations (see
\citet{leq71} and also \citet{thb98} and references therein). Lower resolution images indicate
that there is emission over at least 2\aminn. The shortest spacing available in the D 
configuration of the VLA implies that our images are not sensitive to structures on scales
larger than $\sim$2\aminn. We estimate that the total flux density of NGC 5253 in our image
is 80\% of the single-dish flux density at 8.3 GHz.
The radio continuum emission in NGC 5253 peaks around region
`A' (the cm-emission peak in Turner, Ho and Beck 1998) which contains the two 
main star forming regions - clusters 4 and 5 (CMB97) \nocite{calzetti97}.
The radio continuum emission is weak to the south of the peak, where the 
older clusters are believed to have dispersed the gas.
The radio emission in He 2-10 covers the two main star forming regions - `A' and `B' (region `A' is 
the central 6\asec region and region `B'
is the secondary star forming region 8.5\asec east of `A', \citet{ckv93}). 
Figure 2 shows the \h92 line spectra of the two galaxies at the position of 
the peak continuum emission. The line emission in both the galaxies is 
spatially unresolved, even in the higher resolution C array data.

\section{Modeling the \h92 line} 

In this section, we model the observed data in terms of the local electron density, 
temperature and size of the ionized gas. The models for the ionized gas are 
based on the observed \h92 line flux density and the continuum flux densities
at 8.3 GHz, 4.8 GHz and 15 GHz (see Table 2) from the region of RRL emission. 
Since the line formation process is dependent on the
geometry of the emitting gas, we consider four different
models.  Model I is for a single spherical HII region of fixed size and constant
density. Model II consists of a uniform slab of 
constant density, and fixed thickness and beam filling factor. Model III
consists of a collection of HII regions, of fixed density and size. 
The number of such 
HII regions is normalized using the observed line flux density. 
The third model is similar to those described by \citet{azgv93} and
further adapted by \citet{zagv96} and \citet{pag98} for modeling 
RRLs from ionized gas in starburst regions. 
For all three models, a parameter search is done in density 
\ne in the range 10$^{-2}$ \ccc $<$~n$_e < 10^6$ \ccc and size $l$ in the range 0.01 
pc $<$ $l$ $<$ L pc, where L is the upper limit to the size of the
line emitting region given by the beam size, which is 150 pc for NGC 5253 and 315 pc for He 2-10. 
Model II also includes the beam filling factor as a free parameter, which is 
the lateral extent of the slab of ionized gas within the region of total
lateral size L pc. 
We also consider a fourth model (Model IV) in which the co-existence of multiple densities is
assumed. In this model, we assume a population of HII regions with different densities, but 
with each HII region having a constant density.
We consider those densities which correspond to 
compact (n$_e>~$10$^3$ \ccc and diameter d between 0.1 and 1 pc) and ultracompact HII regions 
(n$_e>~$10$^4$ \ccc and d$~<~$0.1 pc). A HII region of density
\ne is assigned a size $l$ given by the relation \ne= c$_1~l^{\alpha}$. The parameters 
$\alpha$ and c$_1$ depend on the ambient medium and the evolutionary phase of the HII 
region. In our model, these are free parameters. There is evidence for
the existence of such a power law for galactic compact and ultracompact
HII regions (see, for example, Habing and Israel 1979 \nocite{habing79} or Garay 
et al. 1993\nocite{garay93}). 
The population of HII regions of various densities is described by a power-law 
distribution function, N(n$_e$) $\propto$ n$_e^{\beta}$, which is normalized by 
the observed line flux density.
The distribution function for galactic diffuse HII regions is
a power law \citep{kenni89}. \citet{comeron96} found evidence for a power law luminosity
function for galactic ultracompact HII regions, although \citet{cass00} claim 
a peaked distribution function. 
The actual form of the distribution would depend in detail on the 
star formation history of the starburst region as well as on the evolution of 
the HII regions. In the model, we use \b as a free parameter. The fourth free parameter 
in Model IV is n$_e^{min}$, the minimum density in a given distribution. 

The observed data was modeled using models I through IV. Valid solutions are identified as
those combinations of free parameters which produce both the observed \h92 line as well as 
free-free emission at 4.8, 8.3 and 15 GHz less than the corresponding total 
observed continuum flux densities (which would include both thermal and non-thermal 
components). The gas
mass and the ionizing photon flux were also derived for each valid solution. All solutions
were computed for T$_e$ = 10000 and 12500 K. These high temperatures have been derived
from optical spectroscopy and are consistent with the low metallicity of these
two galaxies \citep{camp86,walshr89,kobul97}. Model results for the two 
galaxies are summarized in Tables 3 and 4. 
Since Models I and II give similar results, they are grouped together.
The allowed range of values of densities and sizes of ionized gas are listed. 
The range of densities and sizes allowed by the models are quite limited.
The gas properties for the models with the minimum allowed value of \nlyc are given in the
lower half of Tables 3 and 4. The significance of the minimum allowed \nlyc is as follows :
For an arbitrary continuum optical depth $\tau_c$, 
\nlyc $\propto$ S$_{ff}$ $\times$ $\frac{\tau_c}{1-e^{-\tau_c}}$, where S$_{ff}$ is the
free-free radio continuum flux density. Hence, for an observed value of
S$_{ff}$, the calculated value of the required ionizing photon flux \nlyc increases with increasing
continuum optical depth, $\tau_c$. A given value of S$_{ff}$ thus implies 
a lower limit to the value of 
\nlycc. We take the minimum consistent \nlyc corresponding to the
minimum consistent \tc for modeling. The values for the derived parameters given in Tables 3
and 4 correspond to the average
of the range of values allowed by the n$_e$-$l$ solutions. There are two classes of models
allowed by the existing data : \ne$<~$2500 \cc, $l\sim$5 pc and \ne$>~$2500 \ccc and $l\sim$30 pc
The model solutions are indicated in separate rows for each class
of solutions. The predictions of some
of the valid models for the free-free continuum and line flux densities as a function
of frequency are shown in Figure 3. The calculated values of the free-free continuum flux
densities are less than the observed total continuum flux densities (which are marked as stars in
the figure). The remaining continuum emission which is unaccounted for in our models will 
have contributions both from thermal continuum from any non-RRL emitting gas and also 
from the non-thermal continuum (the latter, only in the case of He 2-10, since NGC 5253 has 
very little contribution from non-thermal emission in the central region). If a significant amount 
of such a non-RRL emitting thermal gas is present,
the amount of such gas will be strongly constrained by the continuum flux densities at $\nu<~$1 GHz.
All the above models will be referred to as `low-resolution' models since these
models are based on data with a resolution of 5\asecc, corresponding to a linear size of 150 pc
for NGC 5253 and 315 pc for He 2-10. In all four models, the favored parameters are
densities of \ne$\sim$10$^4$ \cc, effective sizes of 3-8 pc and masses of ionized gas of
$\sim$10$^4$ \msun. 

Tables 3 and 4 show that for both the galaxies, the four models give results which are 
consistent. But the minimum value of \nlyc varies slightly among the various models. This 
variation is due to the different geometries assumed in the models and
illustrates the uncertainities associated with modeling
data with a limited angular resolution.
In section 4, we discuss the implications of these results of the \lr models.

\subsection{Model IV and the single-density approximation}

As is well known, radio recombination line formation is a non-LTE process. This property can be
used to derive densities from observations if the geometry is known. From the theory
of RRLs, it is known that a particular recombination line is most efficiently emitted from
gas at a particular density and that gas at lower densities emits RRLs preferrentially at lower
frequencies and vice versa \citep{zagv96}.
This property of RRLs (as density filters) implies that a particular recombination
line would receive maximum contributions from a restricted range of gas densities. Hence
previous work which modeled single RRL data using the single-density approximation (corresponding
to Models I,II and III in this work) 
were reasonably successful \citep{azgv93,zagv96,zagv97,pag98,anan00}.
Therefore, with multi-frequency RRL data, the gas can be modeled
as a combination of a minimum number of distinct density components. 
This method was successfully employed to derive a three-density component
model of the ionized gas in Arp 220 based on line and continuum 
data from 1.4 GHz to 230 GHz \citep{anan00}.

However, the single-density approximation based on the above argument is valid only for 
certain functional dependences of emission measure on 
density (i.e., on the relative values of emission measures for gas at different 
densities : the line flux density is most strongly peaked as a function of density when 
$\beta$+2+3/$\alpha$=0. For the line intensity, the condition is $\beta$+2+1/$\alpha$=0. 
See section 3 for definitions of \a and \b). 
For example, Figure 4(a) shows the dependence of the expected line intensity 
at 8.3 GHz as a function of density, for a constant EM and the line strength does
indeed have a peak as a function of electron density. On the other hand,
if a statistically well-sampled population of HII regions with a
power law distribution of EM with density (as in model IV) is assumed, the
line strength no longer peaks for a particular density, but instead, is a monotonic
function of density. For example, it can be seen in Figure 4(b) that when \nlyc is held constant 
(i.e, EM$~\propto~$n$_e^{4/3}$), the line intensity is no longer a peaked function of density.
The results of model IV show that solutions exist only for values of n$_e^{min}$ ranging between
2500 and 25000 \cc. Also, the solutions with the minimum values of \nlyc (the minimum value of
\nlyc which is consistent with the given data, corresponding to the minimum
needed continuum optical depth, see section 3 for an explanation) correspond to
densities between 5000 and 10000 \cc. Though no upper limit to the density is imposed on the models,
the nature of the allowed solutions is such
that only gas of densities very close to n$_e^{min}$ contributes to the line flux density and also
absorbs almost all of the required ionizing UV photons, i.e., all the higher density gas 
in the distribution contributes negligible amounts of RRL emission; this result is common to all valid
solutions in this model. We find that the allowed combinations of the free parameters \a and
l$_{min}$ (the latter being derived from  n$_e^{min}$ and c$_1$ : 
see section 3) are such that the value of \nemin is about 10$^4$ \ccc and the size of the gas
at this density (which is the dominant contributor to the total line strength) is between
3-10 pc. This result explains why
the results of models I, II and III agree well with those of model IV. Hence model IV, 
which is the first attempt at a consistent multi-density model for extragalactic RRL data, 
validates the assumption of a single-density ionized gas in modeling the line emission.  
It should be noted that the narrow range of allowed densities near $\sim$10$^4$ \ccc is
not the density at which the \h92 line emission is expected to peak (from Figure 4(a), this
density is $\sim$10$^2$ \cc). That both the constant density models as well as the multiple
density models give the same density of $\sim$10$^4$ \cc, vastly different from 
the density at which \h92 is expected to peak, is
reason enough to believe that most of the ionized gas in this region does indeed have a 
density close to 10$^4$ \ccc and that the amount of gas at a density of 10$^2$ \ccc must be insignificant
compared to the higher density gas. 

However, the above argument works only if the density distribution is a power law. 
In general in a multi-density environment, where the total EM of gas of a given density is
an arbitrary function of density, an approach based on Model IV will not work. In such a case,
multi-frequency RRL data is essential and models have to be constructed to explain the data
with a minimum number of different density components (as was done in the case of Arp 220 
by Anantharamaiah et al. 2000\nocite{anan00}).

\section{Results from the low-resolution RRL data}

\subsection {NGC 5253}

Most of the models discussed in the previous section yield electron 
densities in excess of a few 1000 \ccc within an effective
size of 2-10 pc (Tables 3 and 4). The minimum value of the derived \nlyc is 
2.5$\times$10$^{52}$ s$^{-1}$. 
This value of \nlyc is computed from the low-resolution data and is 
consistent with the ionizing photon flux computed from
other nebular diagnostics listed below. The continuum flux density of 28 mJy 
measured from the
line emitting region (which is the position of the peak continuum emission, corresponding
to the 6.5\asecc$\times$3.3\asec sized resolution element) in the 
8.3 GHz continuum image (Table 2) corresponds to \nlyc=
3$\times$10$^{52}$ s$^{-1}$, based on the assumption of an optically thin gas. 
The \brg flux from the central 5\asec region \citep{dsw98} implies an \nlyc 
of 1.5$\times$10$^{52}$ s$^{-1}$. The \ha flux density inside an aperture 
of 5\asec (measured using the \ha image of CMB97 \nocite{calzetti97}) gives 
an \nlyc of 1.0$\times$10$^{52}$ s$^{-1}$. 

Taking the average value of \nlyc= 2.5$\times$10$^{52}$ \s and using a Salpeter 
IMF with m$_{upper}$=80 \msun and the tables given in \citet{vgs96}, 
the number of O stars (O3-O9) is estimated to be $\sim$1800. It should be noted that we have used 
the numbers from \citet{vgs96} for solar metallicity whereas the two galaxies NGC 5253 and
He 2-10 have sub-solar metallicities; therefore the calculated values of \nlyc will be
underestimated. However, this error is less than 20\% \citep{vacca94}.
Assuming m$_{lower}$ to be 1.0 \msunn, the total number of stars
within a region of size $\sim$5 pc is $\sim$3.5$\times$10$^5$. This $\sim$5 pc sized region of
gas could be spread over an entire resolution element (150 pc for NGC 5253).
On the other hand, if the gas is physically in a compact region of size $\sim$5 pc,
then the corresponding stellar density indicates that the ionizing sources may well be  
SSCs \citep{meurer95}.

\subsection{Henize 2-10}

Similar to the results for NGC 5253, most models for He 2-10 yield electron densities in excess of 
a few 1000 \ccc within an effective size of 2-10 pc (Table 4). Gas
densities of 1000 \ccc have been derived in the central 2\asec region using 
[SII] line ratios \citep{sugait92}. The minimum derived value of \nlyc 
for the 5\asec sized RRL emitting region is 4$\times$10$^{52}$ s$^{-1}$. The central 2\asec region
has been resolved into a number of star-forming knots and
the value of \nlyc derived from observations which resolve these knots is comparable to
that derived here. \citet{john00} obtain a value of 
\nlyc=~2.5$\times$10$^{52}$ \s from their UV continuum images, while \citet{doyon92} derive a value of  
4$\times$10$^{52}$ \s from their measured \brg flux inside a 5\asec aperture. For an
\nlyc of 4$\times$10$^{52}$ s$^{-1}$, the number of O stars required is $\sim$2900 and the total
stellar mass is $\sim$5.5$\times$10$^5$ \msun (using a Salpeter IMF with m$_{upper}$=80 \msun
and m$_{lower}$ to be 1.0 \msunn, and the results of Vacca, Garmany, \& Shull 1996\nocite{vgs96}). 
As in the case of NGC 5253, the corresponding stellar density for a 3-10 pc region corresponds to an
SSC, if we assume that the ionized gas is physically confined to this $\sim$5 pc region
(see section 5.2 for more details). 

Though the derived ionizing photon flux and the length-scale of the gas for both galaxies imply 
an SSC, the 8.3 GHz RRL data alone does not clarify whether the ionized
gas is in a single clump of diameter $\sim$5 pc or it is spread out in some unknown geometry over the
entire 5\asec region. In the latter case, the ionizing source need not be
an SSC. In the next section, we present arguments that the ionized gas is indeed much
less than 5\asec in size. We also discuss the implications of combining
our RRL data and model results with recently published sub-arcsecond resolution continuum images 
of the centers of the two galaxies (obtained by TBH00 \nocite{tbh00} for NGC 5253 
and by KJ99 \nocite{kj99} for He 2-10). 

\section{Compactness of the RRL emission : evidence from other high resolution observations}

\subsection{NGC 5253} 

We can indirectly infer that the RRL emitting region is smaller than 5\asec from the \ha
and \hb images of CMB97 \nocite{calzetti97}. The \ha image in the inner 5\asec 
is quite clumpy with extensive filaments and knots. Though the RRL emission need 
not follow the \ha emission, it is unlikely that the high density gas responsible for RRL emission 
is spread smoothly over the 150 pc region, given the patchy appearance of the \ha emission. 
and hence we may conclude that the
RRL emitting gas must also be patchy within the 5\asec beam. 

TBH00 \nocite{tbh00} have imaged NGC 5253 at 23 GHz and 15 GHz with the VLA with angular 
resolutions of 0.13\asec
and 0.2\asecc, respectively. A compact ($\sim$0.1\asecc, or 2 pc) source
with a flux density of 11 mJy at 15 GHz, and hence a 
brightness temperature of 12000 $\pm$ 3000 K, was discovered. This was interpreted as a HII
region, and in addition, 
these authors detect radio emission at a low level with a size of 1-2\asecc. 
Further confirmation of the thermal nature of this source comes from recently detected IR
emission from the central object at 11.7 and 18.7 $\mu$m \citep{gorjian01}.
Though the RRL emission is not expected to exactly follow the morphology of continuum emission
(since a particular RRL comes from a limited range of densities, and the continuum 
emission is sensitive to all densities if optically thin), no line emission is 
expected from continuum-free regions and hence we can associate
the observed \h92 emission with the inner 1-2\asec region.
Assuming that the observed radio continuum emission from this galaxy is purely 
due to free-free emission, TBH00 \nocite{tbh00} derive an rms electron density of 
4 $\times$ 10$^4$ \ccc and an \nlyc of 4 $\times$ 10$^{52}$ \s. From the ionization
requirement and the size of the object, TBH00 \nocite{tbh00} suggest that this compact source is a \sn
surrounding an SSC. Since this source is the strongest radio continuum source in 
the inner 5\asec region and since the derived properties of the `supernebula', especially the local
electron density, match those derived from
the RRL data, we tentatively identify the RRL emitting gas with the continuum source detected 
by TBH00 \nocite{tbh00} . Therefore, we have modeled the \sn to estimate the fraction of the observed RRL emission
this compact source is capable of producing. The models are constructed to produce the observed
11 mJy continuum emission at 15 GHz from gas of size $\sim$0.1\asecc. In addition, the continuum
and RRL emission by this gas at 8.3 GHz is also constrained to be less than the observed values 
(the latter two constraints are imposed as upper limits since the observed values correspond to 
the larger 5\asec area). The results of this model are summarized in Table 5. Only 10-35\% of the 
observed 8.3 GHz line emission can be explained by the TBH00 \nocite{tbh00} `supernebula'. All densities
$\geq$ 5$\times$10$^4$ \ccc are consistent with the existing observations. The remaining
line emission presumably originates in the surrounding larger nebulosity with a size of 1-2\asec.
A calculation similar to the one above, where we model the inner 3\asec region with a 15 GHz flux
density of 22 mJy (TBH00)\nocite{tbh00}, shows that such a model is indeed capable of explaining the
entire observed RRL line emission.

The `supernebula', as imaged by TBH00 \nocite{tbh00} is separated by about 0.9\asec from the position of 
the brightest \ha knot in CMB97 \nocite{calzetti97}. This knot was identified with the UV Cluster 5 
in the HST image by \citet{meurer95}. Since the 1$\sigma$
error between the absolute reference frames of the Guide Star 
Catalogue 1 of the HST, on which the \ha and UV images are based, and 
the extragalactic radio reference frame, on which the images in TBH00 \nocite{tbh00} are based, 
is 0.5\asec (Russell et al 1990 \nocite{russell} and also 
Calzetti, and Rajagopal, private communication), the radio \sn of TBH00 \nocite{tbh00} can also be identified 
with the \ha peak (and hence the UV cluster 5) within these errors. The significance of this
identification is two-fold. (1) An additional highly obscured SSC need not be invoked 
and (2) we can relate the known properties of the optical cluster
with those of the `supernebula'. The \ha flux for
this cluster is 4.8$\times$10$^{-16}$ W m$^{-2}$ (see CMB97 \nocite{calzetti97}) 
from which we infer the value
of \nlyc to be $\sim$1.3$\times$10$^{51}$ s$^{-1}$.  The discrepancy between this value and 
\nlyc$>$2.5$\times$10$^{52}$ \s which is derived from
radio continuum data (Table 5) can be attributed to extinction. However, as the age 
of the optical cluster is well constrained, the dynamical age 
of the ionized \sn around it can also be constrained (see section 7.1 for further details).
 
\subsection{Henize 2-10}

The central 5\asec region harbors about nine UV clusters, possibly SSCs, arranged 
in an arc of size $\sim$2\asec \citep{cv94}. This region has a large 
patchy extinction and also clumpy ionized gas. The \brg, UV and optical continuum
images of the center show the presence of several knots \citep{dsw98,cv94,john00}.
KJ99 \nocite{kj99} have imaged the radio
continuum emission of the nucleus of Henize 2-10 with an angular resolution of 0.5\asec using
the VLA. They detect 4-5 compact knots, each of diameter $\sim$0.6\asecc, arranged within a 
5\asecc$\times$2\asec region. These knots of star formation detected individually in the optical, 
\ha, \brg and radio continuum
are not positionally consistent, possibly due to intrinsic astrometric errors.
Attempts at aligning the various images using global astrometric offset corrections
have shown that not all knots have counterparts at all wavelengths (KJ99 \nocite{kj99} 
and Johnson et al. 2000\nocite{john00}); thus there appears to be hidden 
regions of star formation in the 
central 5\asec region. This morphological evidence for the clumpy gas at the center indicates
that the RRL emitting gas is also clumpy on scales $<$5\asec.  

Based  on the continuum spectra, KJ99 \nocite{kj99} infer that the radio knots consist of 
thermal gas at densities 1500-5000 \ccc (optically thick at 8.3 GHz) and also
derive a total \nlyc of 4$\times$10$^{52}$ s$^{-1}$. They suggest that each compact knot may be
ionized by a SSC. We have modeled this region to look for solutions
which produce the observed radio continuum flux densities at 15, 8 and 5 GHz, as reported
by KJ99 \nocite{kj99} . The allowed solutions are also constrained to produce either all or a part 
of the observed \h92 line. The model results are listed in Table 5 and these radio knots,
if thermal, are indeed capable of producing all of the observed \h92 recombination line. 
Therefore we identify these multiple radio continuum knots with the gas responsible for
emitting the observed RRL at 8.3 GHz and associate the ionizing sources with SSCs.

\section{NGC 5253 and Radio-FIR correlation}

The FIR luminosity and the radio luminosity of galaxies which are dominated by 
star formation as opposed to an AGN are very tightly correlated with an rms of
$\leq$0.2 dex on the logarithmic scale \citep{condon92}. This radio-FIR correlation holds for
several types of galaxies such as normal spirals, dwarf starbursts and
dwarf irregulars (see \citet{condon92}). The physical reason for this correlation
is not well understood in detail but may result from a chain of events involving star formation.
The thermal (free-free) portion of the radio emission is from gas ionized by OB stars and the
non-thermal (synchrotron) portion of the radio emission arises from the 
cosmic rays accelerated in SNRs, the end point of Type II SNe caused by massive stars. 
The warm (T $>$ 30K,
S$_{60\mu m}$/S$_{100\mu m}$ $\geq$ 1) dust portion of the FIR emission arises from dust
directly heated by OB stars and the cool (T $<$ 30 K, S$_{60\mu m}$/S$_{100\mu m}$ $<$ 0.8)
dust portion of the FIR emission arises from dust heated by the diffuse interstellar radiation 
field. \citet{xu92} have shown in the case of the LMC that the warm dust emission 
correlates with the
thermal radio emission, while the cool dust emission correlates independently with the 
non-thermal radio emission. Also, the ratio of the two correlated quantities for both these
independent correlations was shown to be the same as that between the the total FIR and the
total radio emission strengths.

The radio continuum of most disk galaxies is dominated by non-thermal emission since
the thermal fraction is typically 0.05-0.2 at 1.4 GHz \citep{condon92}. But it is known that the
BCDGs have a higher fraction of thermal emission than disk galaxies: \citet{kw87} 
showed that the average radio spectral index $\langle\alpha\rangle$ is $-$0.4 for 
BCDGs, while it is $-$0.7 for normal spirals, S$_\nu\propto\nu^\alpha$. Though the mean spectral
index is indeed flatter for BCDGs, the dispersion in the index is much larger than for spirals and
hence they did not estimate an average thermal fraction for the BCD galaxy sample.
BCDGs also have a higher dust temperature (Wunderlich and Klein 1988, \nocite{wk88} 
hereafter WK88) compared
to spiral galaxies which implies that their FIR emission is dominated by emission from warm
dust. It has been suggested by \citet{kw87} and WK88 \nocite{wk88} that the reason the 
BCDGs also obey the radio-FIR correlation is because this dominant warm dust emission correlates
with the thermal radio emission, which is the dominant portion of the radio continuum.

NGC 5253 may well be an unique case which will help to understand the details of the 
radio-FIR correlation. Radio continuum flux density measurements at various frequencies
using single dishes and low resolution images have shown that the radio emission of NGC 5253
is consistent with being almost purely thermal (see \citet{beck96} and also the spectral index
map of Turner, Ho, \& Beck 1998\nocite{thb98}). Its total flux density at 1.4 GHz
is 89 mJy (from the NVSS catalogue of Condon et al. 1998\nocite{condon98} and single-dish observations) 
and the FIR flux is calculated to be 8$\times$10$^8$ \lsun
from the IRAS flux densities. From the ratio of the radio to the FIR luminosity, it can be seen
that this galaxy obeys the radio-FIR correlation \citep{condon92} to within 1.5$\sigma$.
This is a surprising result, since the non-thermal emission in this galaxy is less than 10\% of the
total emission at 5 GHz \citep{beck96,thb98}. 
It is known that the radio flux density and the FIR flux density of galactic HII regions are 
also correlated, but with a different ratio \citep{wwb74,fuerst87,boh89}. On the other hand, 
the ratio of radio to FIR flux densities of compact HII regions is not a constant, but depends 
on a variety of factors \citep{kcw94}. It is worth noting that though the radio emission of
NGC 5253 is almost purely thermal, the radio to FIR emission ratio is similar to star forming 
galaxies rather than HII regions. One way to explain the correlation is
the following : the dust temperature in the nucleus of NGC 5253 is high; \citet{calzetti95}
estimated it to be 51 K. This high dust temperature is the reason why the ratio 
of the strength of the warm dust emission to the cool dust emission is about 16, 
one of the largest among BCDGs and disk galaxies \citep{calzetti95}. Therefore we suggest that
the reason NGC 5253 obeys the radio-FIR correlation is that the warm dust emission and the thermal
radio emission are indeed correlated with each other,
and the slope of the correlation is the same as that between
the total FIR and total radio in normal spirals, as suggested by \citet{xu92} and WK88 \nocite{wk88}. 
WK88 \nocite{wk88} had also put forward an alternative explanation for BCDGs obeying 
the radio-FIR correlation.
They had suggested that a possible lower dust fraction in these galaxies would compensate for the
lower non-thermal radio fraction. This explanation is unlikely for NGC 5253, since among
all BCDGs, it has the lowest non-thermal fraction but the highest dust to gas ratio \citep{lisenf98}.

\section{Discussion}

The detection of a RRL from the central region of both galaxies with parameters similar 
to those derived by TBH00 \nocite{tbh00} and KJ99 \nocite{kj99} 
is added confirmation for the thermal nature of the central compact source(s).
Additionally, an RRL study of dense ionized gas has the following advantages : 
(1) Instead of the rms electron density as given by the continuum data, local electron
densities, independent of filling factor, can be derived from RRL data. (2) Therefore, if the line
emitting region is spatially resolved, then the actual value of \nlyc can be derived (and not
just the lower limit). (3) Due to non-LTE effects, RRLs retain their sensitivity to density even 
in gas which is moderately optically thick, unlike continuum emission. 

\subsection{Dynamics of ionized gas around SSCs}

A question not yet addressed regards the dynamics and time-scales involved in explaining
the existence of $\sim$10$^4$ \msun of ionized gas within a few parsecs of an SSC. In this
section, we will identify the `supernebula(e)' with the RRL emitting 
gas in both the galaxies and also tentatively associate the `supernebula' in NGC 5253 with
the star cluster in the center. The observed
FWHM of $\sim$100 \kms~(estimated from the spectrum before offline hanning smoothing) 
within a 2-8 pc sized region implies a dynamical expansion time
of 2-8$\times$10$^4$ years, quite short compared to the estimated age of 1$-$3
$\times$10$^6$ years for the central cluster in NGC 5253. Therefore, either the supernebula 
must have been set into motion extremely recently, or there must exist a confining mechanism 
operating in the central region. In addition, strong stellar winds from OB and 
Wolf-Rayet stars as well as the natural expansion of the
HII region due to overpressure would drive the ionized gas 
outwards. Given the large mass, high density and small size of the observed HII region ($\sim$2 
pc for the \sn of TBH00 \nocite{tbh00} and KJ99 \nocite{kj99} and 5-8 pc for the 
RRL emitting gas), dynamical 
models should constrain both the age of the SSC as well as the density of the ambient 
unshocked medium. For NGC 5253, for example,
if the \sn is identified with the UV Cluster 5, then the age is determined to be less than
3 million years (CMB97). \nocite{calzetti97}

If we assume the HII region to have reached a diameter of 1 pc solely
due to expansion driven by overpressure in $\sim$2 Myr, then the radius of the initial Stromgren 
sphere would be $\sim$0.01 pc \citep{spitzer}. If at the time of formation of the
initial Stromgren sphere, the ionizing photon flux were $\sim$10$^{52}$ s$^{-1}$, then the corresponding
initial density should be $\sim$2$\times$10$^7$ \cc. In such a case, the mass of the 
shocked neutral gas between the shock and the ionization fronts would be more than 100 times 
greater in the ionized gas inside the ionization front \citep{savgr55}.

In addition, stellar winds will definitely play an important dynamical role. Using 
StarBurst 99 \citep{sb99}, a mechanical luminosity of 10$^{38}$ ergs \s is estimated
for an \nlyc of 10$^{52}$ s$^{-1}$, assuming a continuous starburst model. 
Assuming that the isothermal phase sets in early in the lifetime of the HII region, the 
radius of the thin swept-up ionized shell 
is given by R$_{iso}$ = 16 ($\frac{L}{n_oV}$)$^{1/4}$ t$_6^{1/2}$ pc, where
L is the mechanical luminosity in units of 10$^{36}$ ergs s$^{-1}$, V is the velocity of the stellar
wind at infinity (assumed to be 1000 \kms), n$_o$ is the density of the ambient medium and 
t$_6$ is the time in millions of years \citep{mccray}. 
The thickness of the ionized shell is given by n$_o$R$_{iso}$/(3n$_s$) \citep{dandw}, where  
n$_s$, the density of the shocked ionized gas, is given by n$_s$=n$_o$$\times$V$_s^2$/C$^2$ for an
isothermal shock (V$_s$ is the velocity of the shock, dR$_{iso}$/dt, and C is the speed of
sound in the ambient medium). The values of n$_o$ and t can be expressed in terms of
n$_s$ and the mass of the ionized gas. The range of input parameters are 
10$^3$ \ccc $\leq$ n$_s~\leq$ 5$\times$10$^4$ \ccc and 10$^3$ \msun $\leq$~M $\leq$ 10$^4$ \msun, 
which are similar to the ranges for the models in Tables 3 and 4. The ranges of the derived
values are 0.01 \ccc $\leq$ n$_o$ 
$\leq$ 100 \ccc and 10$^5$ yr $\leq$ t $\leq$ 2$\times$10$^6$ yr. Although the derived values
do not violate any known constraints, the allowed solution space and a detailed
modeling based on future higher resolution continuum images would be useful to understand
the dynamics and the ambient environment of these regions. 
For an initial attempt at such an analysis, see \citet{tankee00}.

\subsection{High density ionized gas in starburst regions}

The commonly occuring relatively high density of $\sim$10$^4$ \ccc for the compact ionized gas 
in the starburst regions in He 2-10 and NGC 5253 and also in other starburst galaxies
(Gilbert et al 2000, Zhao et al 1997, Anantharamaiah et al 2000, TBH00, \nocite{tbh00} 
KJ99 \nocite{kj99}) 
\nocite{gilb00,zagv97,anan00,tbh00,kj99} is striking. One model which may be able 
to explain this commonly occuring high densities is that of \citet{jogdas92,jogdas96}.
In their model, a GMC in the center of a starburst region is shocked
by the overpressure of the surrounding molecular inter-cloud medium (ICM).
The outer shell of the GMC is shock-compressed and moves inwards, and at a 
later stage, due to onset of gravitational instabilities, it 
fragments into star forming centers. This model can well explain the formation of SSCs of mass 
$\geq$10$^5$ \msunn. An estimate of the
density of the shell fragments when star formation begins in the imploding shell is possible.
From \citet{jogdas96}, the velocity of the shock, V$_s$, at the onset of 
gravitational instabilities can
be as high as $\sim$30 \kms and the radius of the shock front reaches $\sim$40\% of the
GMC radius. Adopting a value of the sound speed, C$_s$, inside the GMC as 4 \kms, the 
density of the shocked gas, n$_s$=n$_o~\times$V$_s^2$/C$_s^2$, is increased and the value 
of the density enhancement n$_s$/n$_o$ can
be greater than 55 (this enhancement can be up to three times larger than this value, for
the observed range of values of GMC and ICM properties). In their model, \citet{jogdas92,jogdas96}
assume that the density of the GMC falls
radially as r$^{-1}$ and that the density at the boundary is at least 100 \cc. Therefore this 
density enhancement could explain the observed densities of 10$^3$-10$^4$ \ccc observed in RRLs. 
It remains to be seen if there is indeed a high
pressure reservoir of molecular gas or shocked hot gas at the centers of these galaxies, 
and if so, whether star formation occurs via this process. It should be noted that CO images
of the centers of these two galaxies do not show evidence for gas in this region and therefore
we appeal to whichever gas reservoir gave rise to the SSCs in the center to provide the
neccesary high pressure. However the high density ambient 
medium is not needed as in
the previous cases, since remnant clouds of density $\sim$10$^4$ \ccc would still remain in the
vicinity of the star cluster and be ionized.

The RRL observations in this paper and the recent continuum observations of `supernebulae' may well be
the first detection of ionized gas around extremely young SSCs. Since these clusters are usually too
obscured to be studied even in the IR, higher resolution radio continuum and line data can be used 
to infer the properties of the ambient medium around SSCs, and ultimately, the formation mechanism 
of these massive clusters. Sensitive line data will also yield kinematic information about these
nebulae. But an important problem is the identification of the radio components with the optical and UV
components at the sub-arcsecond level, which should be achievable in a few years.

\section{Summary}

We have detected the \h92 recombination line from two dwarf starbursts NGC 5253 and He 2-10. The
ionized gas responsible for both line emission and the radio continuum emission is modeled in 
terms of electron density and size. The line emitting gas in both galaxies is found to be of
densities $\sim$10$^4$ \ccc with a size of 2-10 pc, implying an \nlyc of a few times 10$^{52}$ s$^{-1}$. 
We also consider a multi-density model for the RRL-emitting gas and show that the assumption 
of a constant density in previous models is valid in a wide range of cases.
Though the line emission is spatially unresolved and the linear size of the beam area corresponds to
150 pc and 315 pc respectively for NGC 5253 and He 2-10, we present arguments which suggest that the
line emission is compact (few parsecs). Based on the derived parameters, these
lines seem to originate from ionized gas associated with SSCs in both galaxies.
The possible `supernebulae' around these clusters have been detected in radio continuum emission 
and we show that these nebulae are capable of producing all the observed line emission in
both the galaxies. 

Both the detection of RRL emission as well as direct radio continuum imaging of these \sn 
around SSCs pose
problems about the lifetimes and dynamics of the ionized gas. We present simple arguments to show that
reconciling the dynamical age of the \sn with the age of the SSCs might be difficult and that detailed 
observational data as well as theoretical modeling are neccesary in order to understand the 
dynamics of HII regions around
these massive clusters. Finally, we point out that though NGC 5253 has a thermal radio spectrum, it
seems to obey the radio-FIR correlation, which is known to hold for galaxies dominated 
by non-thermal radio emission. This feature may well be evidence for the warm 
dust portion of the FIR emission to be
correlated with the thermal radio emission, independent of the cold dust FIR emission and the
non-thermal radio emission, as suggested by \citet{xu92}.

\acknowledgements The National Radio Astronomy Observatory is a facility of the 
National Science Foundation operated under cooperative agreement by Associated 
Universities, Inc. This research has made use of the NASA/IPAC Extragalactic 
Database (NED) which is operated by the Jet Propulsion Laboratory, California 
Institute of Technology, under contract with the National Aeronautics and 
Space Administration. We would like to express our appreciation to the Starburst 99
team for having made their code public and also for making it possible to run the code online.
A part of this work was carried out by NRM and KRA at the National Radio Astronomy 
Observatory in Socorro, New Mexico and they thank NRAO for financial support. 
We thank Daniela Calzetti for the use of her \ha image of NGC 5253, 
Chanda Jog for discussions and Dwarakanath for a careful reading of the 
paper and useful comments. We also thank the referee J.L. Turner for useful
comments and suggestions. This research has made use of NASA's Astrophysics 
Data System Abstract Service.

\begin{figure}
\figurenum{1}
\epsscale{1.1}
\plottwo{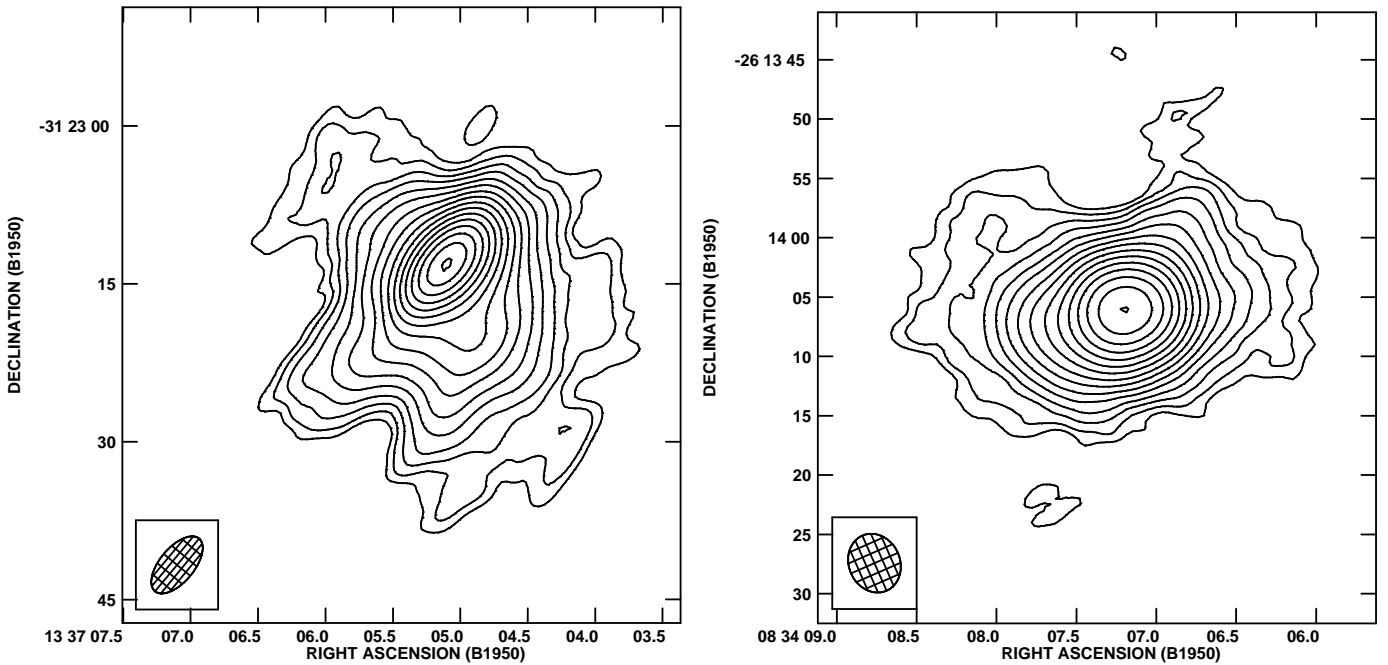}{f2.ps}
\caption{8.3 GHz radio continuum images of NGC 5253 ({\em left}) and He 2-10 ({\em right})
observed using the combined C and D array VLA data, made with `natural weighting' of the 
visibilities. The contours
start at 5$\sigma$ (150$\mu$Jy and 160$\mu$Jy for NGC 5253 and He 2-10 respectively) 
and increase in steps of $\sqrt{2}$. The beam sizes are 6.5\asecc$\times$3.3\asecc, PA=$-$40\deg 
for NGC 5253 and 5.0\asecc$\times$4.3\asecc, PA=24\deg for He 2-10. For further details, 
see Table 2.}
\end{figure}

\begin{figure}
\figurenum{2}
\epsscale{1.1}
\plottwo{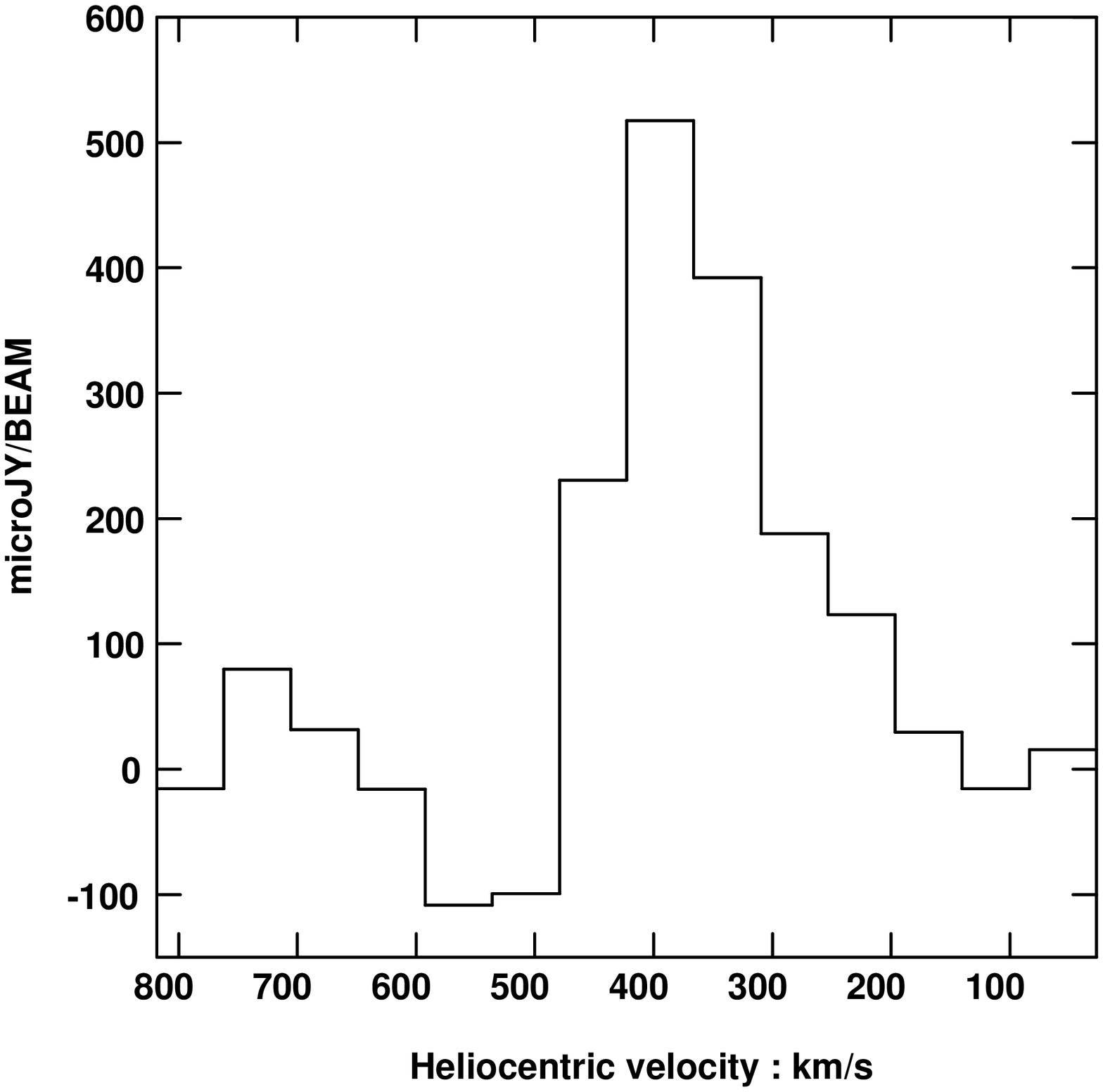}{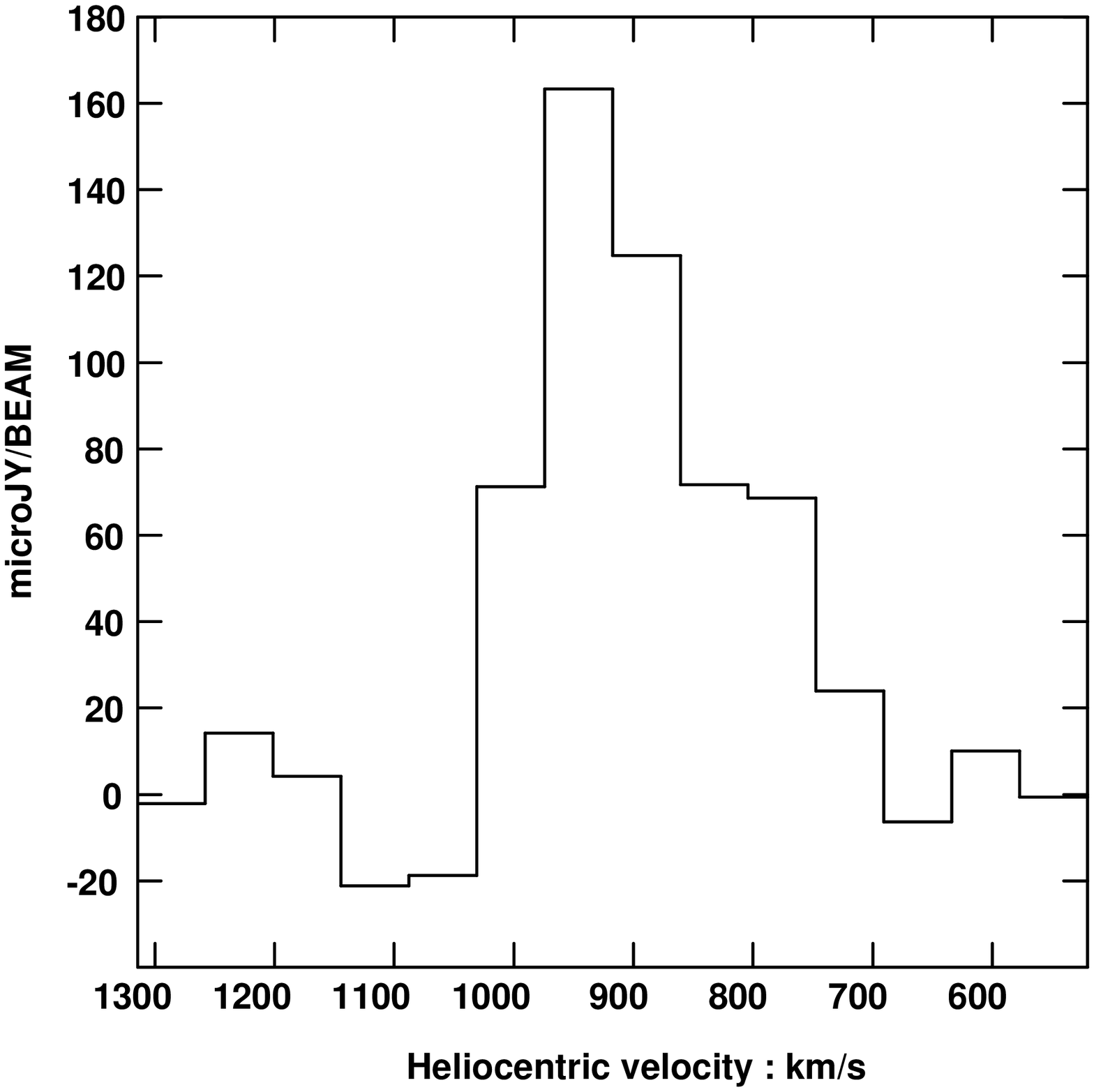}
\caption{The hanning smoothed \h92 spectra at the continuum peaks of NGC 5253 ({\em left}) 
and He 2-10 ({\em right}) at 
8.3 GHz for the combined C and D array VLA data. The line emission is unresolved in
both galaxies. The spatial resolution is 6.5\asecc$\times$3.3\asec for NGC 5253 and
5.0\asecc$\times$4.3\asec for He 2-10. The velocity resolution of the spectrum is 113 \kms.}
\end{figure}

\begin{figure}
\figurenum{3}
\epsscale{0.8}
\plotone{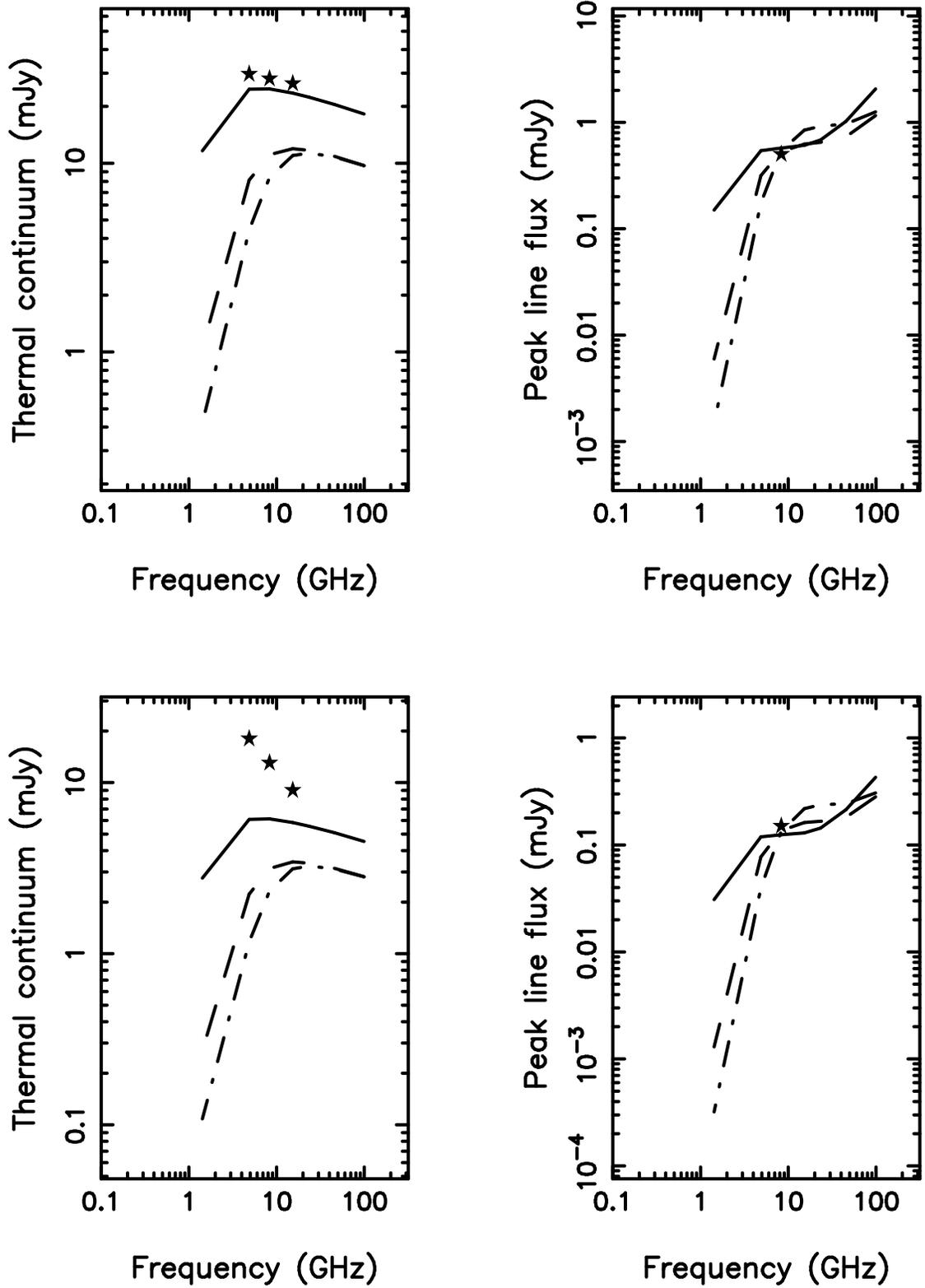}
\caption{Predictions for the thermal continuum and the peak RRL flux densities as a function
of frequency, for NGC 5253 ({\em top}) and He 2-10 ({\em bottom}) for a few valid solutions
corresponding to Model I (see section 3 for details). The data points are marked 
as stars. The solid line
corresponds to n$_e$=10$^3$ \cc, l=24 pc, T$_e$=10000 K, the dashed line to n$_e$=5000 \cc, 
l=6 pc, T$_e$=10000 K and the dot-dashed line to n$_e$=10$^4$ \cc, l=4 pc, T$_e$=10000 K.
The predictions of Models II-IV are similar.}
\end{figure}

\begin{figure}
\figurenum{4}
\epsscale{1.1}
\plottwo{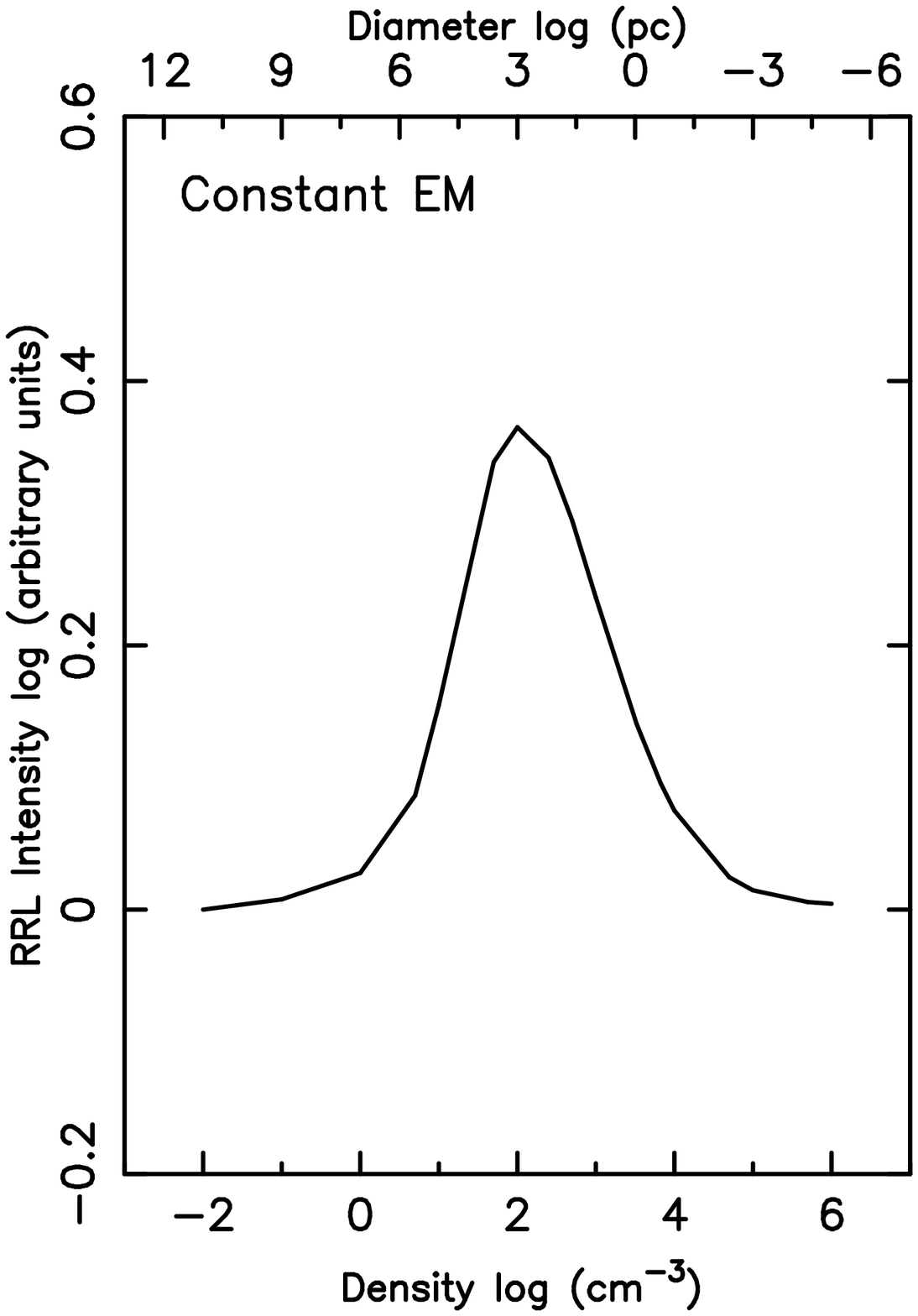}{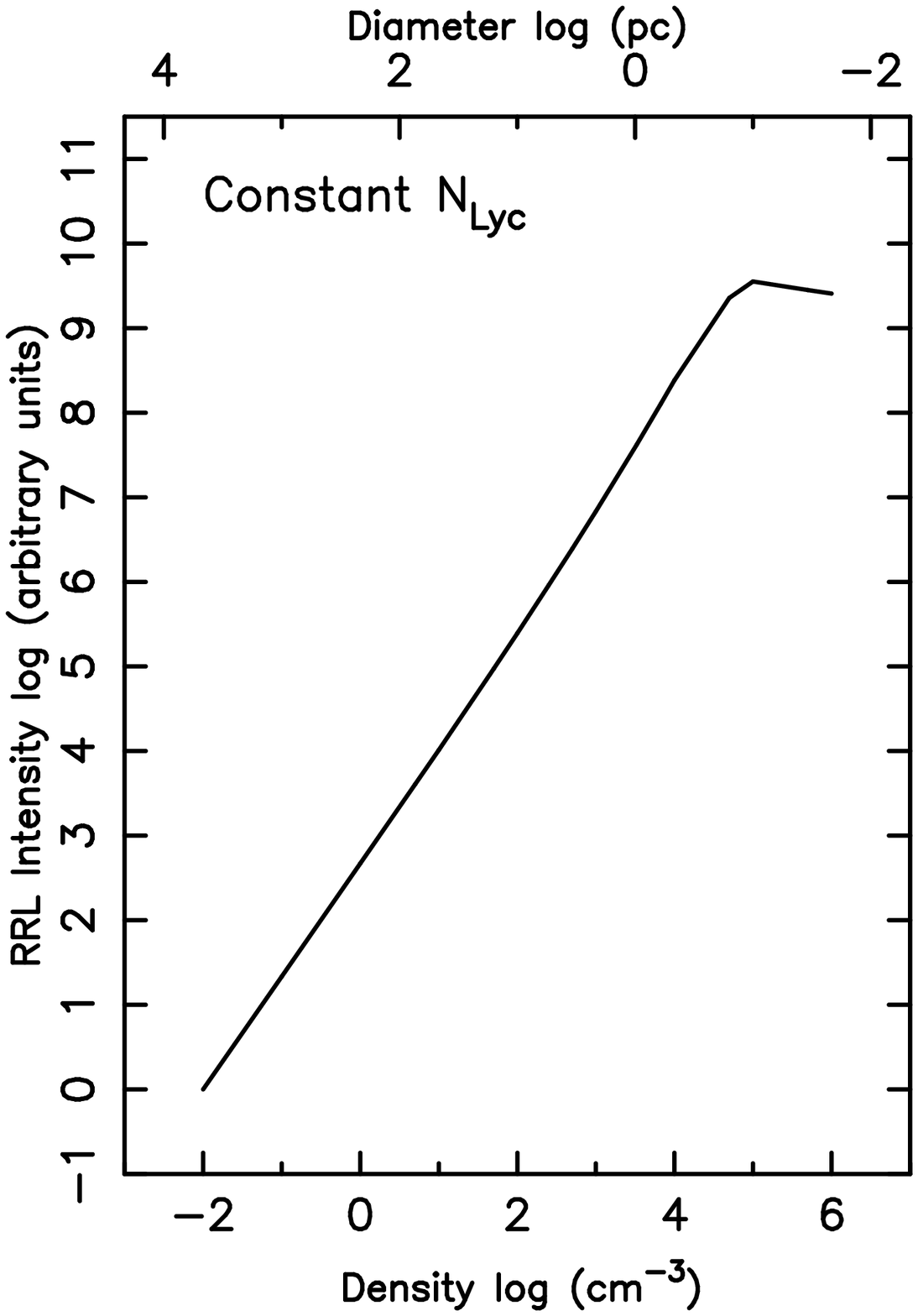}
\caption{Expected \h92 intensity as a function of density for (a) constant EM=10$^7$ \emu 
and (b) constant N$_{Lyc}$=3$\times$10$^{43}$ s$^{-1}$. 
Pressure broadening is neglected and non-LTE effects are taken into account. The 
curve for constant \nlyc illustrates the point
that for an arbitrary population of multi-density ionized gas, the RRL strength 
as a function of density need not a have a maximum over a restricted range of densities.}
\end{figure}

\newpage

\begin{deluxetable}{lcccccc}
\tabletypesize{\small}
\tablecolumns{7} 
\tablewidth{0pc} 
\tablenum{1}
\tablecaption{\bf Log of VLA observations}
\tablehead{ 
\colhead{Galaxy}& \colhead{Array} & \colhead{Date of} &\colhead{uv range} &\colhead{Phase} &\colhead{Bandpass} &\colhead{V$_{hel}$}  \\
\colhead{} & \colhead{} & \colhead{Observation} &\colhead{k$\lambda$} &\colhead{cal} &\colhead{cal} &\colhead{km s$^{-1}$} }
\startdata 
NGC 5253 & DnC & 19 Feb 1999 &0.7-42 & 1334-127 & 1334-127 & 404 \\
NGC 5253 & CnB & 10 Nov 1998&1.3-150 & 1334-127 & 1334-127 & 404 \\
Henize 2-10 & DnC & 19 Feb 1999 &0.7-46  & 0919-260 & 0919-260 & 873 \\
Henize 2-10 & CnB & 10 Nov 1998 & 2.0-130  & 0919-260 & 0919-260 & 873 \\
\enddata 
\end{deluxetable} 

\begin{deluxetable}{lcc}
\tabletypesize{\small}
\tablecolumns{3} 
\tablewidth{0pc} 
\tablenum{2}
\tablecaption{\bf Observational data}
\tablehead{ 
\colhead{Parameter}& \colhead{NGC 5253} & \colhead{He 2-10}}
\startdata 
Beam size & 6.5$^{\prime\prime}\times$3.3$^{\prime\prime}$ & 
5.0$^{\prime\prime}\times$4.3$^{\prime\prime}$\\
P.A. of the beam & $-$40\deg & 24\deg\\
Peak continuum flux  &  &  \\
density at 8.3 GHz\tablenotemark{a}  & 28 mJy & 13 mJy \\
Total continuum flux & &  \\
density at 8.3 GHz\tablenotemark{b}  & 58 mJy & 29 mJy \\
Noise in continuum image (1$\sigma$) & 27 $\mu$Jy & 21 $\mu$Jy \\
Peak H92$\alpha~$line flux density\tablenotemark{c}  & 0.53 mJy & 0.16 mJy \\
Line width (FWHM)  & 95 $\pm$ 10 \kms & 112 $\pm$ 20 \kms\\
(before hanning smoothing) & & \\
Noise in channel image (1$\sigma$)  & 65 $\mu$Jy & 45 $\mu$Jy \\
4.8 GHz flux density & 30 mJy\tablenotemark{d} & 18 mJy\tablenotemark{e} \\
15 GHz flux density & 26 mJy\tablenotemark{d} & 9 mJy\tablenotemark{e}\\
\enddata 
\tablenotetext{a}{Peak continuum flux density in the central 5\asec region}
\tablenotetext{b}{Total continuum flux density measured over the entire C+D continuum
image of 30-45\asec extent : 80\% of the single dish flux density (see section 2)}
\tablenotetext{c}{Line flux density in the central beam area of 5\asec} 
\tablenotetext{d}{Continuum flux density for the central 5\asec : Turner, 
Ho and Beck 1998}
\tablenotetext{e}{Continuum flux density for the central 5\asec, assuming 
S$_\nu$ $\propto$ $\nu^{-0.6}$, see Kobulnicky and Johnson 1999 for a summary 
of the spectral index of He 2-10}
\end{deluxetable} 

\begin{deluxetable}{lccc}
\tabletypesize{\small}
\tablecolumns{4} 
\tablewidth{0pc} 
\tablenum{3}
\tablecaption{\bf The \lr model results for NGC 5253}
\tablehead{ 
\colhead{Parameter}& \colhead{Models I\&II} &\colhead{Model III}&\colhead{Model IV}}
\startdata 
n$_e$ (\cc)\tablenotemark{a}& 500-1000, &  5000-10$^4$ & 2500-25000  \\
           &5000-10$^4$&  &  \\
size (pc)\tablenotemark{a,b} & 20-30, & 5-10 & 3-7  \\
                           & 2-7    &        &  \\
EM (\emu)\tablenotemark{a} & 2$\times$10$^7$ &  10$^7$-10$^8$ & $\sim$10$^8$ \\
          & 2$\times$10$^8$ &       & \\
M$_{HII}$ (\msunn) & 10$^5$,10$^4$ &  10$^4$ & 10$^4$ \\
\cutinhead{Solutions for minimum \nlyc}
N$_{Lyc}^{min}$ (\s) & 1.5$\times$10$^{52}$ & 2.5$\times$10$^{52}$ & 3.5$\times$10$^{52}$ \\
n$_e$ (\cc)  & 5000-10$^4$ & 5000-10$^4$ & 5000-10$^4$  \\
size (pc)\tablenotemark{b} & 2-7 & 4-7  &  3-7  \\
EM (\emu) & 2$\times$10$^8$ & $\sim$10$^8$ & $\sim$10$^8$   \\
S$_{th}$/S$_{obs}^{cont}$ & 0.3-0.5 & $\sim$0.7 & 0.3-0.5  \\
M$_{HII}$ (\msunn) & 10$^4$ & 10$^4$ &  5000-10$^4$  \\
No. O stars\tablenotemark{c} & 1100 & 1825 & 2600  \\
Mass of stars (\msunn)\tablenotemark{d} & 2$\times$10$^5$ & 3.5$\times$10$^5$ & 5$\times$10$^5$  \\
\enddata 
\tablenotetext{a}{~The two rows correspond to two classes of models, both of which are consistent
with existing observations. The models predict different RRL and continuum flux densities at other
frequencies (see Figure 3)}
\tablenotetext{b}{~`Effective' sizes are quoted for Models III and IV, {\em i.e.,} 
N$_{HII}^{1/3}$ $\times~l$, where there are N$_{HII}$ HII regions of size $l$.}
\tablenotetext{c}{O3-O9.}
\tablenotetext{d}{Assuming a Salpeter IMF, m$_{upper}$=80 \msun and 
m$_{lower}$=1.0 \msunn.}
\end{deluxetable} 

\begin{deluxetable}{lccc}
\tabletypesize{\small}
\tablecolumns{4} 
\tablewidth{0pc} 
\tablenum{4}
\tablecaption{\bf The \lr model results for He 2-10}
\tablehead{ 
\colhead{Parameter}& \colhead{Models I\&II} &
\colhead{Model III}&\colhead{Model IV}}
\startdata 
n$_e$ (\cc)\tablenotemark{a}&  500-1000, & 5000-50000 & 2500-25000  \\
           & 5000-10$^4$ &   & \\
size (pc)\tablenotemark{a,b} &  30-40, & 2-10 & 2-10 \\
                           & 3-8     &       &  \\
EM (\emu)\tablenotemark{a} &  2$\times$10$^7$ & 5$\times$10$^7$-2$\times$10$^9$ & $\sim$10$^8$\\
          & $\geq$5$\times$10$^8$ &            & \\
M$_{HII}$ (\msunn) & 3$\times$10$^5$,$\sim$10$^4$ &  7000-40000&10$^4$ \\
\cutinhead{Solutions for minimum \nlyc}
N$_{Lyc}^{min}$ (\s) & 2.5$\times$10$^{52}$ & 4$\times$10$^{52}$ &  5$\times$10$^{52}$\\
n$_e$ (\cc) & 5000-10$^4$ & 5000-50000 & 5000-10$^4$ \\
size (pc)\tablenotemark{b} & 4-7 & 5-10 & 3-7 \\
EM (\emu) &  $\geq$5$\times$10$^8$ & $\sim$10$^8$ & 3$\times$10$^8$ \\
S$_{th}$/S$_{obs}^{cont}$ & 0.2-0.5 & 0.3-0.7 & 0.3-0.6 \\
M$_{HII}$ (\msunn) & $\sim$10$^4$ & 2$\times$10$^4$ & 10$^4$-50000 \\
No. O stars\tablenotemark{c} & 1825 & 2900 & 3650  \\
Mass of stars (\msunn)\tablenotemark{d} & 3.5$\times$10$^5$ & 5.5$\times$10$^5$ & 7$\times$10$^5$ \\
\enddata 
\tablenotetext{a}{~The two rows correspond to two classes of models, both of which are consistent
with existing observations. The models predict different RRL and continuum flux densities at other
frequencies (see Figure 3)}
\tablenotetext{b}{~`Effective' sizes are quoted for Models III and IV, {\em i.e.,} 
N$_{HII}^{1/3}$ $\times~l$, where there are N$_{HII}$ HII regions of size $l$.}
\tablenotetext{c}{O3-O9.}
\tablenotetext{d}{~Assuming a Salpeter IMF, m$_{upper}$=80 \msun and 
m$_{lower}$=1.0 \msunn.}
\end{deluxetable} 

\begin{deluxetable}{lcc}
\tablecolumns{3} 
\tablewidth{0pc} 
\tablenum{5}
\tablecaption{\bf Model results for 8.3 GHz line and the high resolution 
continuum data of KJ99 and TBH00}
\tablehead{ 
\colhead{Parameter}& \colhead{NGC 5253} & \colhead{He 2-10}}
\startdata 
\ne (\cc) & 5$\times$10$^4$-10$^6$ & 5000-10000 \\
size (pc) & 1.0-2 & $\sim$7 \\
\nlyc (\s) & $>$(2.5-4.0)$\times$10$^{52}$\tablenotemark{a} & $>$2.5$\times$10$^{52}$ \\
\tc & $\geq$10& $\sim$0.6 \\
$[$S$_{H92\alpha}]_{model}$/$[$S$_{H92\alpha}]_{obs}$  & 0.1-0.35 & $\sim$1\\
M$_{HII}$ (\msunn) & $\geq$2000 & $\geq$2$\times$10$^4$ \\
No. O stars\tablenotemark{b} & $>$1900-2900  & $>$1900 \\
Mass of stars (\msunn)\tablenotemark{c}&$>$(3.5-5.5)$\times$10$^5$&$>$3.5$\times$10$^5$ \\
\enddata 
\tablenotetext{a}{The range 2.5-4 is over various models.}
\tablenotetext{b}{O3-O9.}
\tablenotetext{c}{~Assuming a Salpeter IMF with m$_{upper}$=80 \msun and
m$_{lower}$=1.0 \msunn.}
\end{deluxetable} 

\end{document}